\input epsf
\documentstyle[preprint,eqsecnum,aps]{revtex}

\begin{document}

\count255=\time\divide\count255 by 60 \xdef\hourmin{\number\count255}
  \multiply\count255 by-60\advance\count255 by\time
  \xdef\hourmin{\hourmin:\ifnum\count255<10 0\fi\the\count255}

\draft
\preprint{\vbox{\hbox{JLAB-THY-00-07}
}}

\title{Large $N_c$, Constituent Quarks, and $N$, $\Delta$ Charge
Radii}

\author{Alfons
J. Buchmann$^\ddagger$\footnote{alfons.buchmann@uni-tuebingen.de} and
Richard F. Lebed$^\S$\footnote{lebed@jlab.org}}

\vskip 0.1in

\address{$^\ddagger$Institute for Theoretical Physics, University of
T\"{u}bingen, D-72076 T\"{u}bingen, Germany\\ $^\S$Jefferson Lab,
12000 Jefferson Avenue, Newport News, VA 23606}

\vskip .1in
\date{March, 2000}
\vskip .1in

\maketitle
\tightenlines
\thispagestyle{empty}

\begin{abstract}
	We show how one may define baryon constituent quarks in a
rigorous manner, given physical assumptions that hold in the
large-$N_c$ limit of QCD.  This constituent picture gives rise to an
operator expansion that has been used to study large-$N_c$ baryon
observables; here we apply it to the case of charge radii of the $N$
and $\Delta$ states, using minimal dynamical assumptions.  For
example, one finds the relation $r_p^2 - r_{\Delta^+}^2 = r_n^2
-r_{\Delta^0}^2$ to be broken only by three-body, $O(1/N_c^2)$ effects
for any $N_c$.
\end{abstract}

\pacs{11.15.Pg, 14.20.-c, 12.39.-x, 13.40.Gp}

\newpage
\setcounter{page}{1}

\section{Introduction}

	The only known path to rendering QCD-like theories
perturbative at all energy scales is to increase the number $N_c$ of
color charges\cite{tHooft}, so that $1/N_c$ itself becomes the small
expansion parameter.  While mesons in large $N_c$ continue to exhibit
the quantum numbers of a single quark-antiquark pair, the large-$N_c$
baryon requires $N_c$ valence quarks, since SU($N_c$) group theory
requires a minimum of $N_c$ fundamental representation indices to form
a color singlet.\footnote{See Ref.~\cite{prague} for a pedagogical
introduction to large $N_c$.}  However, physical baryons consist also
of myriad gluons and sea quark-antiquark pairs; does this then imply
that large $N_c$ baryons have a meaning only within the context of the
valence quark model?  In this paper we claim that this is not the
case, and indeed argue that it is possible to use the very existence
of baryons boasting well-defined quantum numbers and large-$N_c$
arguments to derive a {\em rigorous\/} constituent quark picture.
These assumptions are clearly independent of the momentum transfer
scale, and therefore this constituent picture holds from the
low-energy to deep-inelastic scattering regimes.\footnote{Of course,
for any finite $N_c$, the individual coefficients of the terms in the
$1/N_c$ expansion might grow large for high momentum transfers,
spoiling the utility of the expansion.  It is not known where this
transition occurs.}

	This is actually the same picture, in a somewhat different
language, used to derive an effective Hamiltonian $1/N_c$ operator
expansion for baryon observables.  The operator expansion has been
used to analyze phenomenologically the baryon mass spectrum of the
ground-state\cite{JL}, orbitally-excited\cite{orb}, and
heavy-quark\cite{Jenk2} baryons, as well as magnetic
moments\cite{LMW,DDJM}, axial-vector couplings\cite{DDJM,FJM}, and
photoproduction\cite{CC} and pionic\cite{pion} transitions of $N^*$s
in large $N_c$.

	We then apply this knowledge to a study of the charge radii of
the nonstrange baryons $N$ and $\Delta$.  We first present the generic
expansion demanded by $1/N_c$ when no other physical input is
included, and then specialize to include physical restrictions, such
as the statement that the operators representing the charge radii must
be proportional to the constituent quark charges.  We find that there
are actually two independent contributions at the leading order,
$O(N_c^0)$, and one at $O(1/N_c)$.  Since there are six baryons in the
$N, \Delta$ multiplets, this implies a number of relations between the
charge radii that are expected to be satisfied particularly well, as
we explore below.  For example, we show that a relation found
previously in an $N_c=3$ quark model with two-body currents\cite{BHF}
holds for arbitrary $N_c$ with $O(1/N_c^2)$ corrections.

	The paper is organized as follows.  In Sec.~\ref{const}, we
elucidate the promised relation between constituent quarks and baryon
symmetry properties.  In Sec.~\ref{oper} we restrict to the two-flavor
case and exhibit the complete $1/N_c$ operator expansion for scalar
observables such as $N,\Delta$ charge form factors.  We then consider
this expansion in the ``general parametrization method''\cite{gparm}
generalized to large $N_c$, which places additional restrictions on
the allowed operators based on the observable at hand.  We present and
discuss results in Sec.~\ref{results} and conclude in
Sec.~\ref{concl}.

\section{Large $N_c$ and Constituent Quarks}
\label{const}

	We begin with the quantum numbers of the current quarks
themselves.  To obtain the electric charge and hypercharge of the
quarks for arbitrary $N_c$, we require only that $(u,d)$, $(c,s)$, and
$(t,b)$ remain weak isospin doublets with $I_3 = +1/2$ and $-1/2$,
respectively, that under strong isospin the up quark and down quark
still form a doublet with $I_3 = +1/2$ and $-1/2$, respectively, while
the strange and all other quarks are isosinglets, and that all quarks
in the electroweak interaction and $u,d,s$ quarks in the strong
interaction satisfy the Gell-Mann--Nishijima condition
\begin{equation}
Q = I_3 + Y/2 .
\end{equation}
Then the cancellation of the SU($N_c$)$\times$SU(2)$\times$U(1)
standard model chiral anomalies imposes
\begin{equation}
Q_{u,c,t} = (N_c+1)/2N_c, \ \ Q_{d,s,b} = (-N_c+1)/2N_c, 
\end{equation}
while under strong hypercharge one finds
\begin{equation}
Y_u = Y_d = 1/N_c, \ \ Y_s = (-N_c+1)/N_c .
\end{equation}
It is interesting to note that these results maintain for arbitrary
$N_c$ the usual electric charge and hypercharge assignments familiar
in $N_c =3$, such as the proton quantum numbers $Q_p = Y_p = +1$.

	Baryons in large $N_c$ have masses of $O(N_c)$, owing to both
the intrinsic $O(1)$ masses of the quarks and interaction terms which
also scale as $N_c$\cite{witten}.  The emergence of an exact
spin-flavor symmetry in the large-$N_c$ limit for any number of
flavors was first demonstrated in Ref.~\cite{DM}, so that it is
meaningful to classify baryons into spin-flavor representations at
leading order in $1/N_c$.

	The ground-state multiplet of baryons for arbitrary $N_c$
fills, by assumption, a spin-flavor multiplet described by a tensor
completely symmetric on $N_c$ indices (Fig.~1).  For three flavors
($u,d,s$), this is an SU(6) multiplet that for $N_c=3$ reduces to the
familiar positive-parity ${\bf 56}$-plet containing the spin-1/2 SU(3)
octet and spin-3/2 decuplet.  When $N_c > 3$, these multiplets are
much larger.\footnote{The multiplets are exhibited in
Refs.~\cite{prague,JL,DJM}.} Then each multiplet possesses, in
general, a number of states whose quantum numbers reduce to those of
the familiar baryons in $N_c=3$.  For example, the spin-flavor
multiplet of Fig.~1 decomposes into $N_c$ distinct flavor multiplets
with spins 1/2, 3/2, \ldots, $N_c/2$: Is the $\Delta$ to be identified
as a spin-3/2 or spin-$N_c/2$ state?  In this case, one finds
that\cite{Jenk1} $(M_\Delta - M_N) \propto J(J+1)/N_c$, compared to
$M_{\Delta,N} =O(N_c)$.  The observed relatively small $\Delta$-$N$
mass splitting suggests that one should take $J=3/2$ rather than
$J=N_c/2$.

	Similar considerations\cite{prague} lead one to take the
large-$N_c$ analogues of the familiar baryons to have the usual spins,
isospins, and hypercharges of $O(1)$ rather than $O(N_c)$.  In
particular, this identifies the proton as a state with $I=I_3=1/2$,
$J=1/2$, and valence quark content consisting of the usual triple of
$uud$ in an $I=J=1/2$ combination, augmented by $(N_c-3)/2$ $ud$
pairs, each in a spin-singlet, isosinglet combination.  Then $N_u =
(N_c+1)/2$ and $N_d = (N_c-1)/2$, and one may verify the previous
claim that $Q_p = Y_p = +1$.

	Obtaining a rigorous constituent picture for baryons requires
that each baryon truly resides in a unique spin-flavor multiplet.
In the case of the familiar SU(3) octet and decuplet baryons, this is
the completely symmetric {\bf 56}-plet of SU(6).  Such an assumption
is subject to two conditions:
\begin{enumerate}
\item The baryons are stable under strong interactions, so that they
are true narrow-width eigenstates of the strong Hamiltonian.  This is
true in large $N_c$, since the production of each meson costs one
power of $1/\sqrt{N_c}$ in the amplitude.  It is also true for
physical nucleons, where only weak decays are permitted, and to a
lesser extent for the other ground-state baryons, where phase space
suppresses such decays.

\item Configuration mixing between the dominant ground-state multiplet
and higher multiplets is suppressed.\footnote{``Configuration mixing''
has two meanings here: One, such as that used in the text, indicates
the change of a baryon wavefunction when spins or flavors of
individual quarks are altered.  There is also a narrower meaning of
mixing between two spin-flavor eigenstates with the same global
quantum numbers, such as between nucleon and Roper states.  In both
cases, the mixing between pure spin-flavor eigenstates requires gluon
exchanges and thus is suppressed in $1/N_c$.}  This is also true in
large $N_c$, where such mixing requires the exchange of gluons to
excite the ground state into an overlap with the higher state.  These
gluon couplings introduce additional $1/N_c$ suppressions.  For
example, consider flipping the spin of one of the $N_c$ quarks in a
proton to form a $\Delta^+$.  Dynamics tells us that the spatial
wavefunction of the baryon should adjust itself to the new spin
configuration; however, since only one of the quarks in $N_c$ has
changed, one expects this effect to be suppressed by some power of
$1/N_c$.

\end{enumerate}

	Once these conditions are satisfied, it becomes a matter of
mathematics alone to identify individual ``constituent'' quarks within
the baryon.  This is seen from the spin-flavor Young tableau for the
ground state (Fig.~1); the spin-flavor wavefunction is a completely
symmetric tensor with $N_c$ indices, represented by $N_c$ boxes in the
tableau.  Each index corresponds to a fundamental representation of
the spin-flavor group, and carries precisely the same quantum numbers
as one of the current quarks within the baryon.  One may use
spin-flavor projection operators to isolate these representation
``quarks'' (which we call {\it r-quarks\/}), the collective action of
which is to reproduce the {\em entire\/} baryon spin-flavor
wavefunction.\footnote{The spatial wavefunction of each r-quark then
has the same functional behavior as the spatial wavefunction of the
whole baryon, restating the assumption that configuration mixing is
neglected.}  In terms of field theory, the r-quarks are interpolating
fields carrying the spin-flavor quantum numbers of current quarks,
such that an appropriately symmetrized set of $N_c$ boast complete
overlap with the baryon wavefunction (Fig.~\ref{merc}).  The r-quarks
are then true ``constituent'' quarks, in that the baryon is
constituted entirely of them and nothing else.  To reiterate, {\em the
rigorous constituent quark is the r-quark, which is defined as the
interpolating field associated with a single box in the baryon Young
tableau.}  It turns out that the ``Naive quark model for an arbitrary
number of colors'' presented in Ref.~\cite{KP}, based on the
constituent quark model, is not so naive after all.

	We hasten to add that this is not a revolutionary idea.  It
was understood, at least implicitly, in a number of large-$N_c$
analyses where knowledge of the completeness of sets of spin-flavor
operators acting upon particular baryon multiplets is important, such
as in Refs.~\cite{JL,orb,Jenk2,DDJM,FJM,CC}.  Indeed, the ``quark
representation'' presented in Ref.~\cite{DJM} is mathematically
equivalent to the r-quark construction.  Our purpose in introducing
the r-quark is to give such analyses a firm physical interpretation as
well as to probe the limitations of this picture, as detailed above.

	Obviously, such a manipulation cannot possibly tell us
everything about the baryon structure.  As an explicit example,
consider the strangeness content of the proton.  We have argued that
the flavor structure of the proton for arbitrary $N_c$ consists of the
usual valence $uud$ triple and $(N_c-3)/2$ $ud$ pairs each in a
spin-singlet, isosinglet combination.  But if these are all of the
r-quarks, how can the proton have strange content?  The answer is that
$s\bar s$ pairs are present, as are other sea quarks and gluons, but
all of these have been incorporated into the r-quarks.  In terms of
field theory, these other components have been integrated out in favor
of the r-quark fields.\footnote{Indeed, $s\bar s$ pairs in the proton
appear only in vacuum loops, which introduce $1/N_c$ suppressions
compared to pure glue interactions.  The same is not necessarily true
for $u\bar u$ or $d\bar d$ pairs in the proton, which can appear in
$Z$-graphs.}  Thus, the proton may have strange content even if it has
no strange r-quarks.

	The r-quark decomposition clearly does not indicate in detail
how constituent quarks are formed from the fundamental degrees of
freedom in the baryon.  But it does give, by construction, values for
observable matrix elements that an arbitrarily good constituent quark
model, i.e., one that gives all of the correct baryon observables,
must satisfy.  In this way it serves as a means to improve explicit
quark model calculations.  As an example given in the next section,
one can extract the r-quark masses and interaction energy terms from
the $N,\Delta$ spectrum.

	Finally, it should be pointed out that this decomposition has
nothing to do with large $N_c$ {\it per se}, except that identifying
physical baryons with distinct irreducible spin-flavor representations
for larger $N_c$ is on somewhat more solid theoretical ground because
of the conditions listed above.  If one declares that the proton lies
entirely in an SU(6) {\bf 56}-plet in the physical case of $N_c=3$,
there is no problem in defining the three r-quarks.

\section{Operator Analyses}
\label{oper}

	The analysis of any observable with given spin-flavor quantum
numbers in the $1/N_c$ expansion may be carried out in essentially the
same way: One simply writes down all operators with the same
spin-flavor transformation properties as the observable, weighted with
the appropriate suppression power of $1/N_c$.  The number of such
operators is finite since the number of spin-flavor structures
connecting the initial- and final-state baryons is finite.  As a
trivial example, consider the problem of mass operators of the $I=1/2$
nucleon states.  The Wigner-Eckart theorem tells us that only
operators with isospins $I = 0$ or 1 can connect the states.  Indeed,
the most general decomposition, as was done for the ground-state
baryon masses\cite{JL}, or the orbitally-excited baryons\cite{orb}, or
here for the charge radii, may be considered the application of the
Wigner-Eckart theorem in spin-flavor space.

	We also see from this example that there are precisely as many
operators (2) as independent mass observables, which permits arbitrary
masses for the $p$ and $n$ states.  In the given example, the $I=0$
and $I=1$ operators contribute to $(m_n + m_p)/2$ and $(m_n - m_p)$,
respectively.  Unless some of the operators in a given expansion may
be eliminated or suppressed, the operators merely provide a
reparametrization of the data, i.e., a different basis for the same
vector space of observables.

	However, we have not yet taken into account suppressions of
operators by powers of $1/N_c$.  In order to identify these
suppressions for baryons, it is most convenient to work with r-quarks.
Let us define an {\it n-body operator\/} as one that requires the
participation of $n$ r-quarks; that is, the Feynman diagram has a
piece that is $n$-particle irreducible.  Since r-quarks each carry a
fundamental color index, they exchange gluons just like current quarks
and hence obey the same large-$N_c$ counting rules.  Indeed, it is not
difficult to see that an $n$-body operator requires the exchange of a
minimum of $n-1$ gluons and hence a suppression of $1/N_c^{n-1}$,
since\cite{tHooft} $\alpha_s \propto 1/N_c$.

	The most general possible $n$-body operators can be built from
$n$th-degree polynomials in 1-body operators, whose members fill the
adjoint representation of the spin-flavor group.  We denote these
\begin{eqnarray}
J^i & \equiv & q^\dagger_\alpha \left( \frac{\sigma^i}{2} \otimes
\openone \right) q^\alpha , \nonumber \\
T^a & \equiv & q^\dagger_\alpha \left( \openone \otimes
\frac{\lambda^a}{2} \right) q^\alpha , \nonumber \\
G^{ia} & \equiv & q^\dagger_\alpha \left( \frac{\sigma^i}{2} \otimes
\frac{\lambda^a}{2} \right) q^\alpha , \label{basis}
\end{eqnarray}
where $\sigma^i$ are the usual Pauli spin matrices, $\lambda^a$
denote Gell-Mann flavor matrices, and the index $\alpha$ sums over
all $N_c$ quark lines in the baryons.  In the two-flavor case
considered here, $T^a$ is replaced with the isospin operator $I^a$.
One then builds polynomials in $J$, $I$, and $G$ with the same
spin-flavor quantum numbers as the observable in question.

	However, there are still three important points to take into
account before the analysis is complete.  First, the operators in
Eq.~(\ref{basis}) sum over all the r-quarks in the baryon and may add
coherently to give combinatoric powers of $N_c$ that compensate some
of the $1/N_c$ suppressions.  Generally, this occurs for $G$ and not
$I$ or $J$, since we have chosen baryons to have spins and isospins of
$O(1)$ rather than $O(N_c)$.

	Second, there exist relations, called operator reduction
rules\cite{orb,DJM}, between some combinations of operators due to the
spin-flavor symmetry or the symmetry of the baryon representation.
For example, one particular combination of $J^2$, $I^2$, and $G^2$ is
the quadratic Casimir of the spin-flavor algebra, and just gives the
same number when applied to all baryons in the same representation.
In the two-flavor case with scalar operators, the operator reduction
rules of Ref.~\cite{DJM} tell us that the $G^{ia}$ never need appear,
since every possible contraction of its spin index leads to a
reducible combination.  Likewise, $I^2 = J^2$ in the two-flavor case.

	Third, the most complicated operator necessary to describe a
baryon with $N_c$ r-quarks is an $N_c$-body operator.  However,
ultimately we are interested in the subset of these baryons that
persist when $N_c=3$, and by the same logic, these are completely
described by expanding only out to 3-body operators.  The 4-, 5-,
\ldots, $N_c$-body operators would be linearly independent when acting
upon the {\em full\/} baryon representation, but must be linearly
dependent on the 0-, 1-, 2-, and 3-body operators when acting upon the
baryons that persist for $N_c=3$.  Since we are not taking the strict
$N_c \to \infty$ limit but rather $N_c$ large and finite, the question
of losing information due to noncommutativity with the chiral
limit\cite{Cohen} does not arise.

	Using these rules, it is straightforward to write down the
expansion for an arbitrary scalar quantity with possible isospin
breaking but preserving $I_3$ (as in electromagnetic interactions or
masses).  Our example is the derivative of the baryon charge (Sachs)
form factor $F(q^2)$ at $q^2=0$, but note that the same expansion
would hold for the whole $q^2$-dependent form factor, as well as
masses\cite{JL}:
\begin{equation} \label{matel}
-6 \left. \frac{dF(q^2)}{dq^2} \right|_{q^2 = 0} = \left< c_0 \openone
 + c_1 I_3 + \frac{c_2}{N_c} I^2 + \frac{c_3}{N_c} \left\{ I_3, I_3
 \right\} + \frac{c_4}{N_c^2} \left\{ I^2 , I_3 \right\} +
 \frac{c_5}{N_c^2} \left\{ I_3, \left\{ I_3, I_3 \right\} \right\}
 \right> .
\end{equation}
The brackets indicate that the operators are to be evaluated for a
particular baryon state; anticommutators are used to remind one that
the commutator combinations are reducible, owing to the spin-flavor
symmetry.  Here, each of the coefficients $c_i$ possesses a $1/N_c$
expansion starting at order $N_c^0$; they play the role of reduced
matrix elements in the Wigner-Eckart theorem.  We make the naturalness
assumption that any dimensionless coefficient appearing in the
analysis is of order unity, unless one can think of a reason it is
suppressed (additional symmetry or chance dynamical cancellation) or
enhanced (additional dynamics).  An example of the first case is the
neutron-proton mass difference, where one would find an anomalously
small coefficient unless the approximate symmetry of isospin is
recognized.  An example of the second case is that the neutron-proton
scattering lengths are much larger than ``natural size,'' pointing to
shallow bound (the deuteron) or nearly bound (${}^1 S_0$) states.

	To illustrate the r-quark picture for the baryons, consider
the case of $N$ and $\Delta$ masses using the r.h.s.\ of
Eq.~(\ref{matel}).  The operator $\openone$ clearly gives a common
mass $c_0$ to each r-quark, while the $I_3$ term differentiates $u$
and $d$ r-quarks.  The remaining operators require interactions of the
r-quarks and may be considered matrix elements of the potential. Using
Breit-Wigner masses for the $\Delta$ states (Note, however,
Ref.~\cite{CW} for a treatment using pole masses), one finds
\begin{equation}
m_u = c_0 + c_1/2 = 287.6 \ {\rm MeV}, \ \ m_d = c_0 - c_1/2 = 289.6 \
{\rm MeV} ,
\end{equation}
and the interaction energy terms for nucleons and $\Delta$'s amount to
about 73 and 366 MeV, respectively.  These values for the quark masses
are consistent with those used in ordinary constituent quark models.
The r-quark masses thus account for the bulk of baryon masses,
underscoring the economy of this picture.

	It is convenient to rewrite Eq.~(\ref{matel}) in terms of the
global quantum numbers $J(J+1)$ and $Q$, which equal $I(I+1)$ and $I_3
+1/2$, respectively, in the two-flavor case.  Then the expansion reads
\begin{equation} \label{opoldN}
-6 \left. \frac{dF(q^2)}{dq^2} \right|_{q^2 = 0} = d_0 N_c + d_1 Q +
 \frac{d_2}{N_c} J(J+1) + \frac{d_3}{N_c} Q^2 + \frac{d_4}{N_c^2} Q
 J(J+1) + \frac{d_5}{N_c^2} Q^3 ,
\end{equation}
where again each $d_i$ possesses a $1/N_c$ expansion starting at order
$N_c^0$.  Note in either case that there are 6 independent operators,
reflecting that there are 6 observables, corresponding to the
isodoublet of $N$'s and the isoquartet of $\Delta$'s.  Equation
(\ref{opoldN}) is therefore the most general expansion one can write
down, modified only by the $1/N_c$ suppression factors.

	For the particular case of the charge form factor, one can go
a bit further.  Despite their $O(N_c)$ masses, baryons in large $N_c$
nevertheless have a finite size\cite{witten}, so $d_0 N_c$ in
Eq.~(\ref{opoldN}) should actually be replaced by $d_0$.  One can see
this by noting that no interaction diagram in the baryon is larger
than $N_c^1$, so that the interaction energy per quark is no larger
than $N_c^0$, and thus the wavefunction of each quark has a spatial
extent of $O(N_c^0)$. Thus, the most general expansion based solely
upon symmetry and the grossest features of large $N_c$ reads
\begin{equation} \label{opold}
-6 \left. \frac{dF(q^2)}{dq^2} \right|_{q^2 = 0} = d_0 + d_1 Q +
 \frac{d_2}{N_c} J(J+1) + \frac{d_3}{N_c} Q^2 + \frac{d_4}{N_c^2} Q
 J(J+1) + \frac{d_5}{N_c^2} Q^3 .
\end{equation}

	This operator method lies at one extreme end of possible
analyses, in that it includes {\em only\/} symmetry information.  At
the other end lie phenomenological models, in which not only the
structure of the individual operators but also their coefficients are
provided.  As an intermediate choice, one may impose mild physical
constraints on the allowed operators; this is the approach of the
``general parametrization (GP) method''\cite{gparm}.  It was applied
to the case of baryon charge radii\cite{DiM} in order to check
relations appearing in a quark model calculation\cite{BHF} that
includes two-body exchange currents.  Here we extend the analysis to
arbitrary $N_c$.  Pion-baryon couplings are studied using the GP and
compared with results of a $1/N_c$ approach in Ref.~\cite{BH}.

	It should be stressed that these ``mild physical constraints''
do indeed impose some model dependence on the GP, meaning that its
predictivity follows not from QCD alone but requires additional
dynamical assumptions.  However, as argued next and in the first
paragraph of Sec.~\ref{results}, these assumptions have a firm
dynamical basis and are more mild than those of an arbitrary model.

	The assumptions of the GP method for charge form factors are
quite minimal: All scalar operators are allowed that couple to the
quarks (r-quarks in our case) through precisely one factor of the
quark charges, which is what one expects from a single photon vertex.
Then one has
\begin{equation} \label{opers}
-6 \left. \frac{dF(q^2)}{dq^2} \right|_{q^2 = 0} = A \sum_{i}^{N_c}
Q_i + \frac{B}{N_c} \sum_{i \neq j}^{N_c} Q_i \, \mbox{\boldmath
$\sigma$}_i \cdot \mbox{\boldmath $\sigma$}_j + \frac{C}{N_c^2}
\sum_{i \neq j \neq k}^{N_c} Q_i \, \mbox{\boldmath $\sigma$}_j \cdot
\mbox{\boldmath $\sigma$}_k .
\end{equation}
The rules for assigning $1/N_c$ suppressions in the coefficients are
the same as above: $n$-body operators have a factor $1/N_c^{n-1}$, and
$A,B,C$ each possess $1/N_c$ expansions starting at order
$N_c^0$. Note that this expression, unlike Eq.~(5) in Ref.~\cite{DiM},
has no strange quark term: As discussed above, the $N$'s and
$\Delta$'s have no strange r-quarks; the $s\bar s$ contributions
appear as $O(1/N_c)$ corrections to the dynamical coefficients already
presented.

	It is straightforward to evaluate matrix elements of these
three operators.  The sums are re-expressed in terms of the Casimirs
$Q$, ${\bf J}^2$, ${\bf S}_u^2$, and ${\bf S}_d^2$.  To evaluate the
final two Casimirs, note that the spin-flavor wavefunction is
completely symmetric.  Thus, all of the $u$ quarks, for example, are
in a symmetric state, and one then has total $u$-quark spin $S_u =
N_u/2$.  After simplifying all terms, one finds
\begin{eqnarray}
\sum_i Q_i & = & Q, \nonumber \\
\sum_{i \neq j} Q_i \, \mbox{\boldmath $\sigma$}_i \cdot
\mbox{\boldmath $\sigma$}_j & = & Q (N_c-1) - \left[ N_c + 2(J+1)
\right] \left[ N_c - 2J \right] / 2N_c , \nonumber \\
\sum_{i \neq j \neq k} Q_i \, \mbox{\boldmath $\sigma$}_j \cdot
\mbox{\boldmath $\sigma$}_k & = & Q \left[ 4J(J+1) + 2 - 5N_c \right]
+ \left[ N_c + 2(J+1) \right] \left[ N_c - 2J \right] / N_c .
\end{eqnarray}
The GP expansion then reads
\begin{eqnarray} \label{opnew}
-6 \left. \frac{dF(q^2)}{dq^2} \right|_{q^2 = 0} & = & A Q +
\frac{B}{N_c^2} \left\{ Q N_c (N_c-1) - \frac 1 2 \left[ N_c + 2(J+1)
\right] \left[ N_c - 2J \right] \right\} \nonumber \\ & & +
\frac{C}{N_c^3} \left\{ Q N_c \left[ 4J(J+1) + 2 - 5N_c \right] +
\left[ N_c + 2(J+1) \right] \left[ N_c - 2J \right] \right\} .
\end{eqnarray}
The charge radii, defined as
\begin{equation} 
r^2_B = -6 \left. \frac{1}{F(q^2)} \frac{dF(q^2)}{dq^2} \right|_{q^2 =
0} = -6 \left. \frac{1}{Q} \frac{dF(q^2)}{dq^2} \right|_{q^2 = 0}
\end{equation}
if $Q \neq 0$, and neglecting the $Q$ factor if $Q = 0$, are presented
for the $N$ and $\Delta$ states in Table~I.

	It is interesting to compare the two expressions
Eqs.~(\ref{opold}) and (\ref{opnew}).  First, one sees that the latter
is, as it must be, a special case of the most general possible
expression Eq.~(\ref{opold}).  Specifically, the two expressions are
related by
\begin{eqnarray}
d_0 & = & -\frac B 2 - \frac{1}{N_c} (B-C) + \frac{2C}{N_c^2} ,
\nonumber \\
d_1 & = & A + B - \frac{1}{N_c} (B+5C) + \frac{2C}{N_c^2} ,
\nonumber \\
d_2 & = & \frac{2B}{N_c} - \frac{4C}{N_c^2} , \nonumber \\
d_3 & = & 0 , \nonumber \\
d_4 & = & 4C , \nonumber \\
d_5 & = & 0 ,
\end{eqnarray}
meaning that in GP the coefficients $d_0, d_1, d_4$ are independent
and of natural [$O(1)$] size, $d_2$ is dependent and subleading in
$1/N_c$, and $d_3 = d_5 = 0$.  Note also that the coefficient $B$ can
appear at $O(1)$ and $C$ at $O(1/N_c$), a factor $N_c$ larger than
naively expected from Eq.~(\ref{opers}), a result arising from the
combined spin (\mbox{\boldmath $\sigma$}) and flavor ($Q_i$) structure
of the corresponding operators.  Since the $Q$ operator, containing a
piece transforming as $I=1$, is the sole source of isospin breaking in
the GP, one expects that the $I=2$ and 3 contributions, first
appearing in $Q^2$ and $Q^3$ terms, are absent.  By the Wigner-Eckart
theorem, one can see that these relations involve only $\Delta$
states.

\section{Results and Discussion}
\label{results}

	We have pointed out that the GP expression Eq.~(\ref{opnew})
is not the most general possible expansion for the charge radius.  The
other terms in Eq.~(\ref{opold}) but not (\ref{opnew}) can appear if
subleading effects are taken into account.  For example, in the GP
expression, the only source of isospin quantum numbers is the quark
charge operator $Q_i$.  Explicit isospin breaking due to, say, the
$u$-$d$ quark mass difference introduces factors of the operator $I_3
= Q-1/2$, which do not conform to the expression (\ref{opnew}), but
appear with an additional small ($\sim 5 \times 10^{-3}$) coefficient.
Similar statements are expected for loop corrections; for example, one
can see how electromagnetic loop corrections induce a $Q^3$ and
possibly other suppressed terms in the expansion, at the cost of an
$\alpha_{\rm EM}/4\pi$ suppression.  Inasmuch as these additional
effects are dynamically suppressed, the GP expansion should give an
excellent expansion for the charge form factors.  Since the neglected
coefficients are small, they would make little numerical difference if
included in the analysis below.

	One interesting feature of the GP expression Eq.~(\ref{opnew})
is that the terms not proportional to the total baryon charge $Q$ are
all proportional to $N_c -2J$, and in particular, vanish for
$J=N_c/2$.  That is, all charge radii (and other electromagnetic
matrix elements) are proportional to $Q$ for $J=N_c/2$, which was
pointed out by Coleman\cite{cole} for the case $N_c=3$.  The symmetry
reason for this feature is not hard to see: The charge operator $Q$
transforms according to the adjoint representation of the spin-flavor
group.  The $J=N_c/2$ flavor representation, unlike that of $J=1/2$,
3/2, \ldots, $N_c/2-1$, is completely symmetric, and has the same
Young tableau as the spin-flavor representation in Fig.~1.  In the
product of this representation with its conjugate (relevant to baryon
bilinears) there is only one adjoint representation, and since one
already has one such operator, $Q$, its matrix elements must be
proportional to the eigenvalue $Q$.  For the flavor representations
with $J<N_c/2$ (such as that of spin-3/2 for $N_c>3$), the
corresponding product has two or more adjoints, and exact
proportionality to $Q$ no longer holds.

	As discussed above, $I=2$ and 3 terms are absent in
Eq.~(\ref{opnew}).  The following relations (or any combination
thereof) hold in the GP:
\begin{eqnarray} \label{I23}
2 r_{\Delta^{++}}^2 - r_{\Delta^+}^2 - r_{\Delta^0}^2 -r_{\Delta^-}^2
= 0 & & \ (I=2) , \nonumber \\
2 r_{\Delta^{++}}^2 - 3r_{\Delta^+}^2 +3r_{\Delta^0}^2 +
r_{\Delta^-}^2 = 0 & & \ (I=3) .
\end{eqnarray}

	One also sees from Eq.~(\ref{opnew}) and Table~I that both $A$
and $B$ terms are of leading order ($N_c^0$) for generic $N$'s and
$\Delta$'s in large $N_c$, despite the fact that the former comes from
one-body and the latter from two-body operators.  This is due to the
coherence effect in the two-body operator.  Similarly, the three-body
operator ($C$ term) is suppressed only by $1/N_c$.  It is only
special combinations of the charge radii in which these leading
effects cancel.  A particularly interesting combination of this type
is
\begin{equation} \label{newreln}
(r_p^2 - r_{\Delta^+}^2) - (r_n^2 -r_{\Delta^0}^2) = -12C/N_c^2 ,
\end{equation}
in which the full one- and two-body terms, as well as the coherent
part of the three-body term, cancel for all $N_c$.  This cancellation
also holds for the completely generic expansion (\ref{opold}), in
which the r.h.s. of Eq.~(\ref{newreln}) reads $-3d_4/N_c^2$.  Thus, if
three-body operators are neglected, one has
\begin{equation}
r_p^2 - r_{\Delta^+}^2 = r_n^2 -r_{\Delta^0}^2 ,
\end{equation}
for all $N_c$.  The only other such combinations are obtained by
adding linear combinations of Eqs.~(\ref{I23}).  If we still demand an
$O(1/N_c^2)$ combination but allow a two-body operator (which would
serve to distinguish large $N_c$ from the straightforward GP
approach), one finds the separate relations
\begin{eqnarray} \label{deltas}
r_p^2 - r_{\Delta^+}^2 & = & -\frac{6}{N_c^2} \left[ B + 2C \left( 1 -
\frac{1}{N_c} \right) \right] , \nonumber \\
r_n^2 - r_{\Delta^0}^2 & = & -\frac{6}{N_c^2} \left[ B -
\frac{2C}{N_c} \right] .
\end{eqnarray}
One may combine these relations with Eqs.~(\ref{I23}) to predict all
the $\Delta$ charged radii in terms of $r_{p,n}^2$ good to
$O(1/N_c^2)$.

	Alternately, if one allows $N_c$-dependent coefficients, the
only relation in addition to Eqs.~(\ref{I23}) with no corrections in
the GP is
\begin{equation}
(N_c+5)(N_c-3) \, r_n^2 = (N_c+3)(N_c-1) \, r_{\Delta^0}^2 ,
\end{equation}
which is trivial for $N_c=3$.

	For completeness, the isovector and isoscalar charge radii are
given by
\begin{eqnarray}
r_{I=1}^2 = (r_p^2 - r_n^2) & = & A + B \, \frac{N_c-1}{N_c} - 5C
\, \frac{N_c-1}{N_c^2} , \nonumber \\
r_{I=0}^2 = (r_p^2 + r_n^2) & = & A - 3B \, \frac{N_c-1}{N_c^2} -3C
\, \frac{(N_c-1)(N_c-2)}{N_c^3} .
\end{eqnarray}
The experimental values $r_p^2 = 0.792(24)$ fm$^2$\cite{prot} and
$r_n^2 = -0.113(3)(4)$ fm$^2$\cite{neut}, together with Table~I,
suggest that $A/B \approx 5$ if $C$ is neglected.  While this is
somewhat larger than one would expect from a pure naturalness
assumption, dynamical models for $A$ and $B$ must be studied to decide
whether this ratio is unnatural.  Moreover, using Eqs.~(\ref{I23}) and
(\ref{deltas}) with these experimental values and estimating
$O(1/N_c^2)$ terms to be about $r_p^2/9 \approx 0.09$ fm$^2$ (which
overwhelms statistical uncertainties on $r_{p,n}^2$), one finds
\begin{eqnarray}
r_{\Delta^{++}}^2 & = & r_p^2 - \frac 1 2 r_n^2 + \frac{3}{N_c^2}
\left[ B + 2C \left( 2 - \frac{1}{N_c} \right) \right] = 0.85 \pm 0.09
\, {\rm fm}^2, \nonumber \\
r_{\Delta^+}^2 & = & r_p^2 + \frac{6}{N_c^2} \left[ B + 2C \left( 1 -
\frac{1}{N_c} \right) \right] = 0.79 \pm 0.09 \, {\rm fm}^2, \nonumber
\\
r_{\Delta^0}^2 & = & r_n^2 + \frac{6}{N_c^2} \left[ B - \frac{2C}{N_c}
\right] = -0.11 \pm 0.09 \, {\rm fm}^2, \nonumber \\
r_{\Delta^-}^2 & = & r_p^2 - 2 r_n^2 - \frac{6}{N_c^2} \left[ B - 2C
\left( 1+ \frac{1}{N_c} \right) \right] = 1.02 \pm 0.09 \, {\rm fm}^2.
\end{eqnarray}

\section{Conclusions}
\label{concl}

	We have seen that a rigorous constituent quark picture for
baryons, in that all spin-flavor matrix elements are reproduced by
construction, follows from the assumption that the physical baryons
are narrow-width eigenstates of distinct spin-flavor representations.
Both of these requirements hold in the large-$N_c$ limit.  To improve
systematically upon these assumptions, baryon strong decay amplitudes
and configuration mixing must be accommodated, opening up new
possibilities for large-$N_c$ quark models.

	The analysis of observables is possible in this simplified
scheme.  In particular, here we have studied $N,\Delta$ charge radii,
and showed 1) that the one-body and part of the two-body operator are
of leading order in $1/N_c$, and 2) that a number of useful relations
follow from a simple parametrization (GP) representing the most
important physical effects.  It will be interesting to test which of
these relations are supported by experiment.

{\samepage
\begin{center}
{\bf Acknowledgments}
\end{center}
AJB thanks W. Broniowski for useful discussions and the Deutsche
Forschungsgemeinschaft for support under title BU 813/2-1 and
Jefferson Lab for their hospitality.  RFL thanks C.E. Carlson,
C.D. Carone, and J.L. Goity for valuable comments, and the Department
of Energy for support under Contract No.\ DE-AC05-84ER40150.}

\begin{table}
\begin{tabular}{lrclc}
$r_p^2$ & $A + B \ \frac{\textstyle (N_c-1)(N_c-3)}{\textstyle 2N_c^2}
- C \ \frac{\textstyle (N_c-1)(4N_c-3)}{\textstyle N_c^3}$ & $A -
\frac 2 3 C$ \\
$r_n^2$ & $-B \ \frac{\textstyle (N_c-1)(N_c+3)}{\textstyle 2N_c^2} +
C \ \frac{\textstyle (N_c-1)(N_c+3)}{\textstyle N_c^3}$ & $-\frac 2 3
B + \frac 4 9 C$ \\
$r_{\Delta^{++}}^2$ & $A + B \ \frac{\textstyle
3(N_c^2-2N_c+5)}{\textstyle 4N_c^2} - C \ \frac{\textstyle
3(3N_c^2-12N_c+5)}{\textstyle 2N_c^3}$ & $A + \frac 2 3 B + \frac 2 9
C$ \\
$r_{\Delta^+}^2$ & $A + B \ \frac{\textstyle N_c^2-4N_c+15}{\textstyle
2N_c^2} - C \ \frac{\textstyle (N_c-1)(4N_c-15)}{\textstyle N_c^3}$ &
$A + \frac 2 3 B + \frac 2 9 C$ \\
$r_{\Delta^0}^2$ & $-B \ \frac{\textstyle (N_c-3)(N_c+5)}{\textstyle
2N_c^2} + C \ \frac{\textstyle (N_c-3)(N_c+5)}{\textstyle N_c^3}$ & 0
\\
$r_{\Delta^-}^2$ & $A + B \ \frac{\textstyle 3(N_c^2-5)}{\textstyle
2N_c^2} - C \ \frac{\textstyle 3(2N_c^2-5N_c-5)}{\textstyle N_c^3}$ &
$A + \frac 2 3 B + \frac 2 9 C$
\end{tabular}
\caption{Charge radii of $N$ and $\Delta$ states as functions of
$N_c$ and for $N_c=3$.}
\label{chgrad}
\end{table}

\begin{figure}
  \begin{raggedright}

\def\ssqr#1#2{{\vbox{\hrule height #2pt
      \hbox{\vrule width #2pt height#1pt \kern#1pt\vrule width #2pt}
      \hrule height #2pt}\kern- #2pt}}
\def\sqr{\mathchoice\ssqr8{.4}\ssqr8{.4}\ssqr{5}{.3}\ssqr{4}{.3}}

\def\bsqr{\ssqr{15}{.3}}

\def\nbox{\hbox{$\bsqr\bsqr\bsqr\bsqr\raise7pt\hbox{$\,\cdot\cdot\cdot
\cdot\cdot\,$}\bsqr\bsqr\bsqr$}}
\bigskip

\centerline{$$\nbox$$\mbox{\hspace{2em}}}
\bigskip\bigskip

\caption{The completely symmetric spin-flavor $N_c$-box Young tableau,
corresponding to ground-state baryons.}
\label{young}

  \end{raggedright}
\end{figure}

\begin{figure}
  \begin{centering}
  \def\epsfsize#1#2{0.75#2}
  \hfil\epsfbox{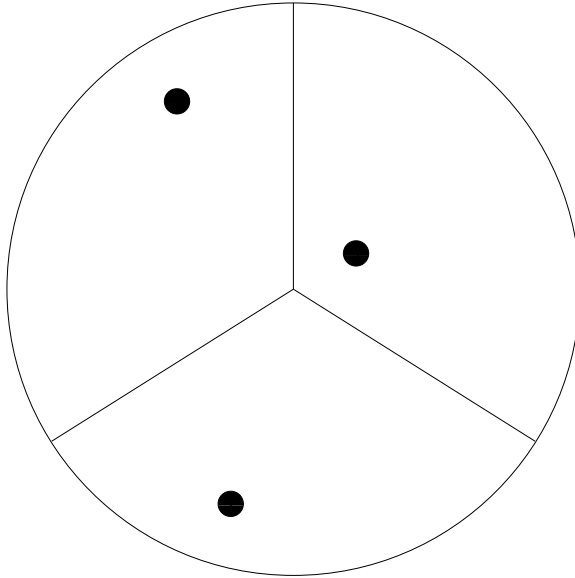}\hfil\hfill

\caption{Qualititative illustration of current quarks (dots)
versus r-quarks (wedges) for $N_c=3$ baryons.  Note that the actual
division is in spin-flavor, not spatial, coordinates.  The entire
baryon, including glue, sea quarks, etc., is subsumed into the
r-quarks.}

\label{merc}

\end{centering}
\end{figure}

\end{document}